\PassOptionsToPackage{unicode}{hyperref}
\PassOptionsToPackage{hyphens}{url}
\PassOptionsToPackage{dvipsnames,svgnames,x11names}{xcolor}
\documentclass[
  letterpaper,
  DIV=11,
  numbers=noendperiod]{scrartcl}
\newcounter{csnum}

\usepackage{xcolor}
\usepackage{amsmath,amssymb}
\usepackage{iftex}
\ifPDFTeX
  \usepackage[T1]{fontenc}
  \usepackage[utf8]{inputenc}
  \usepackage{textcomp} 
\else 
  \usepackage{unicode-math} 
  \defaultfontfeatures{Scale=MatchLowercase}
  \defaultfontfeatures[\rmfamily]{Ligatures=TeX,Scale=1}
\fi
\usepackage{lmodern}
\ifPDFTeX\else
\fi
\IfFileExists{upquote.sty}{\usepackage{upquote}}{}
\IfFileExists{microtype.sty}{
  \usepackage[]{microtype}
  \UseMicrotypeSet[protrusion]{basicmath} 
}{}
\makeatletter
\@ifundefined{KOMAClassName}{
  \IfFileExists{parskip.sty}{%
    \usepackage{parskip}
  }{
    \setlength{\parindent}{0pt}
    \setlength{\parskip}{6pt plus 2pt minus 1pt}}
}{
  \KOMAoptions{parskip=half}}
\makeatother
\makeatletter
\ifx\paragraph\undefined\else
  \let\oldparagraph\paragraph
  \renewcommand{\paragraph}{
    \@ifstar
      \xxxParagraphStar
      \xxxParagraphNoStar
  }
  \newcommand{\xxxParagraphStar}[1]{\oldparagraph*{#1}\mbox{}}
  \newcommand{\xxxParagraphNoStar}[1]{\oldparagraph{#1}\mbox{}}
\fi
\ifx\subparagraph\undefined\else
  \let\oldsubparagraph\subparagraph
  \renewcommand{\subparagraph}{
    \@ifstar
      \xxxSubParagraphStar
      \xxxSubParagraphNoStar
  }
  \newcommand{\xxxSubParagraphStar}[1]{\oldsubparagraph*{#1}\mbox{}}
  \newcommand{\xxxSubParagraphNoStar}[1]{\oldsubparagraph{#1}\mbox{}}
\fi
\makeatother

\usepackage{longtable,booktabs,array}
\usepackage{calc} 
\usepackage{etoolbox}
\makeatletter
\patchcmd\longtable{\par}{\if@noskipsec\mbox{}\fi\par}{}{}
\makeatother
\IfFileExists{footnotehyper.sty}{\usepackage{footnotehyper}}{\usepackage{footnote}}
\makesavenoteenv{longtable}
\usepackage{graphicx}
\makeatletter
\newsavebox\pandoc@box
\newcommand*\pandocbounded[1]{
  \sbox\pandoc@box{#1}%
  \Gscale@div\@tempa{\textheight}{\dimexpr\ht\pandoc@box+\dp\pandoc@box\relax}%
  \Gscale@div\@tempb{\linewidth}{\wd\pandoc@box}%
  \ifdim\@tempb\p@<\@tempa\p@\let\@tempa\@tempb\fi
  \ifdim\@tempa\p@<\p@\scalebox{\@tempa}{\usebox\pandoc@box}%
  \else\usebox{\pandoc@box}%
  \fi%
}
\def\fps@figure{htbp}
\makeatother

\NewDocumentCommand\citeproctext{}{}

\makeatletter
 \let\@cite@ofmt\@firstofone
 \def\@biblabel#1{}
 \def\@cite#1#2{{#1\if@tempswa , #2\fi}}
\makeatother
\newlength{\cslhangindent}
\newlength{\csllabelwidth}
\newenvironment{CSLReferences}[2] 
 {\begin{list}{}{%
  \setlength{\itemindent}{0pt}
  \setlength{\leftmargin}{0pt}
  \setlength{\parsep}{0pt}
  \ifodd #1
   \setlength{\leftmargin}{\cslhangindent}
   \setlength{\itemindent}{-1\cslhangindent}
  \fi
  \setlength{\itemsep}{#2\baselineskip}}}
 {\end{list}}
\usepackage{calc}

\KOMAoption{captions}{tableheading}
\makeatletter
\@ifpackageloaded{caption}{}{\usepackage{caption}}
\AtBeginDocument{%
\ifdefined\contentsname
  \renewcommand*\contentsname{Table of contents}
\else
  \newcommand\contentsname{Table of contents}
\fi
\ifdefined\listfigurename
  \renewcommand*\listfigurename{List of Figures}
\else
  \newcommand\listfigurename{List of Figures}
\fi
\ifdefined\listtablename
  \renewcommand*\listtablename{List of Tables}
\else
  \newcommand\listtablename{List of Tables}
\fi
\ifdefined\figurename
  \renewcommand*\figurename{Figure}
\else
  \newcommand\figurename{Figure}
\fi
\ifdefined\tablename
  \renewcommand*\tablename{Table}
\else
  \newcommand\tablename{Table}
\fi
}
\@ifpackageloaded{float}{}{\usepackage{float}}
\floatstyle{ruled}
\@ifundefined{c@chapter}{\newfloat{codelisting}{h}{lop}}{\newfloat{codelisting}{h}{lop}[chapter]}
\floatname{codelisting}{Listing}

\makeatother
\makeatletter
\@ifpackageloaded{caption}{}{\usepackage{caption}}
\@ifpackageloaded{subcaption}{}{\usepackage{subcaption}}
\makeatother
\usepackage{bookmark}
\IfFileExists{xurl.sty}{\usepackage{xurl}}{} 
\hypersetup{
  pdftitle={Jagged AI in Scientific Peer Review: Evidence from POMP Data Analysis},
  colorlinks=true,
  linkcolor={blue},
  filecolor={Maroon},
  citecolor={Blue},
  urlcolor={Blue},
  pdfcreator={LaTeX via pandoc}}

\title{Jagged AI in Scientific Peer Review: Evidence from POMP Data
Analysis}
\author{Jin Wook Lee, William Szegda, Zhisheng Song, Edward L. Ionides}
\date{\normalsize University of Michigan, Ann Arbor. Correspondence to ionides@umich.edu}
\begin{document}
\maketitle

\subsection*{Abstract}\label{abstract}
\addcontentsline{toc}{subsection}{Abstract}

Despite their growing use in academic writing and statistical analysis,
the performance of artificial intelligence (AI) tools in scientific peer
review remains a largely unexplored area. A key challenge is
\emph{jagged AI}, a phenomenon where AI exhibits strong ability spikes
in some domains while remaining deficient in others. To study this
jaggedness in a practical data science context, we considered the task
of reviewing partially observed Markov process (POMP) data analyses.
POMP models, a generalization of state-space models or hidden Markov
models, are used to fit mechanistic dynamic models to time series data
in diverse applications including disease transmission, ecological
dynamics, and financial risk assessment. High-quality peer review in
this area entails assessment of scientific context, identification of
errors in implementing complex algorithms, and decisions concerning
methodological best practices. We studied 72 POMP projects from four
semesters of a University of Michigan graduate time series course for
which the project reports, the source code, and student peer reviews are
anonymized and open access. We compared the human reviews with four AI
review agents, using Claude Code with differing instructions implemented
as skill files. We found that the AI review agents exhibited a jagged
capability profile, identifying plausible technical errors and instances
of invalid inference methodology overlooked by humans, while showing
lower capability at finding issues involving interpretive errors,
narrative coherence, and domain-informed model critique. The jaggedness
was similar for all agents, consistent with other research finding that
the specific instructions can have limited effect on the capability of
advanced AI models for some tasks. Skill file configuration shifted
which weaknesses agents emphasized, without removing the jaggedness.

\textbf{Keywords:} Peer review; Partially observed Markov process;
Jagged artificial intelligence.

\section{Introduction}\label{introduction}

The use of large language models (LLMs) in academic contexts has grown
substantially, raising important questions about where AI assistance
enhances scientific outcomes and where it degrades them (Vaccaro et al.,
2024). A foundational framework for understanding this boundary is the
concept of the \emph{jagged technological frontier} (Dell'Acqua et al.,
2026). We define \emph{jaggedness} to be a phenomenon where performance
is super-human at some aspects of a task, and sub-human in others. This
normative definition was proposed by (Morris et al., 2026), and the
reference humans for our investigation are those in our study. The term
\emph{jagged} was introduced by (Dell'Acqua et al., 2023), subsequently
published in 2026. Jaggedness has become commonly used to refer to
cognitive gaps in otherwise capable AI systems (Mineault et al., 2026).
Tasks that appear equally demanding to human knowledge workers can fall
on opposite sides of this frontier: where AI capabilities align with the
task, AI complements and amplifies human performance; where they do not,
AI output becomes unreliable and can actively degrade the quality of
work. The benefit of AI assistance is therefore contingent on whether a
given task lies within the current capability boundary---a boundary that
is neither smooth nor predictable. Jaggedness is not a uniform
capability ceiling but a pattern of uneven performance across domains.
Morris et al. (2026) characterize it as a tendency for AI to exhibit
sharp ability spikes in certain areas while remaining persistently
deficient in others, arguing that this unevenness fundamentally
complicates any simple notion of AI generality.

Data science peer review represents a complex, multi-layered task that
is susceptible to jagged AI performance. A reviewer must verify
statistical correctness, evaluate model assumptions, critique scientific
reasoning, assess presentation quality, and apply domain
knowledge---capabilities that vary considerably in how well they lend
themselves to LLM evaluation. Early empirical studies have begun to map
where LLM reviewing succeeds and fails: Liu \& Shah (2023) found that
GPT-4 detects deliberately inserted errors in fewer than half of tested
manuscripts, while ({Liang et al.}, 2024) showed that AI-generated
feedback on papers from major journals overlaps with human reviewers at
rates comparable to inter-reviewer agreement, yet tends toward generic
rather than targeted critique. These studies focus on AI reviewing of
general scientific manuscripts; the performance of AI on technical,
domain-specific statistical analyses, where correctness depends on both
code implementation and inference methodology, remains largely
unexplored. The task of reviewing Partially Observed Markov Process
(POMP) data analysis King et al. (2016) provides a demanding scenario,
requiring specialized knowledge of likelihood-based inference, particle
filter diagnostics, computationally intensive algorithms, and the
interplay between mechanistic model structure and statistical
identifiability.

In this paper, we assess the jaggedness of AI in peer reviewing POMP
data analysis projects, asking two related questions: (1) Which
categories of weaknesses can AI effectively identify in POMP analyses,
and which does it consistently miss? (2) Can this jaggedness be tuned
through the use of specialized instructions, templates, and scripts for
specific workflows? We evaluated four Claude Code agents (i.e.,
instances of Claude Code provided with specific skill files) across 72
open-access POMP projects, comparing their reviews against graduate
student peer reviews validated by the course instructor. The resulting
evidence provides a detailed characterization of AI jaggedness in a
rigorous statistical peer review context, demonstrating that AI can
effectively complement, but not replace, human judgment in scientific
evaluation.

\section{Materials and Methods}\label{materials-and-methods}

We used a corpus of 72 POMP analysis projects from the STATS 531 course
``Modeling and Analysis of Time Series Data'' at the University of
Michigan, spanning four semesters: Winter 2021
(\href{https://github.com/ionides/531w21/tree/main/final_project}{W21,
16 projects}), Winter 2022
(\href{https://github.com/ionides/531w22/tree/main/final_project}{W22,
23 projects}), Winter 2024
(\href{https://github.com/ionides/531w24/tree/main/final_project}{W24,
16 projects}), and Winter 2025
(\href{https://github.com/ionides/531w25/tree/main/final_project}{W25,
17 projects}). The course was not offered in Winter 2023. For each final
project, a group of 2-3 graduate students chose a novel scientific
scenario relevant to POMP data analysis, found and analyzed data, and
wrote a reproducible report. All project reports are available online,
anonymously, together with source code. Each project was subjected to
blind peer review by approximately 5 students. The students had access
to the report and source code. The instructor evaluated each student
review, added their own feedback, and posted a final aggregated review
online. Student reviews were graded on their ability to identify project
weaknesses, but the grades and individual reviews are not publicly
available.

We developed four AI agents using Anthropic's Claude framework, each
constructed as a specialized AI assistant with a custom system prompt
and specific tool access (Anthropic, 2025). All reviews were generated
using Claude Sonnet 4.6 during April--May 2026. All agents had no
internet access or persistent memory, ensuring that each review was
conducted independently without cross-contamination between projects.
Agents were given Bash, file search, and read tool access, allowing them
to search, read, and execute code within the project folder. Each agent
was tasked with constructing a structured peer review for every project,
producing a list of ``major'' and ``minor'' weaknesses as output. Agents
were asked to make no more than 15 points, to be comparable to the human
lists that typically had 10-12 points. This cap ensures that the agent
cannot obtain a high Human Overlap score by generating a large output
volume. The four agents differed in their skill file configuration,
summarized below. The skill files, additional details of the
experimental procedures, and the results obtained, are available at
\url{https://github.com/ionides/jagged-ai-peer-review}.

\textbf{Baseline}: A baseline agent with no access to skill files,
representing the inherent jaggedness of Claude's base model in POMP peer
review.

\textbf{CourseGuided}: This agent was equipped with the
\texttt{guided-pomp-review} skill file, which provided a 13-item
checklist grounded in the advice of Wheeler et al. (2024) for generating
rigorous, evidence-based peer reviews of POMP manuscripts, covering
likelihood inference, benchmark comparisons, convergence diagnostics,
and related evaluation criteria. It additionally had access to a
\texttt{531-references} skill, compiled from past STATS 531 course
materials, providing documented course conventions and a catalogue of
common errors students were explicitly taught to avoid.

\textbf{MetaSkill}: This agent used the \texttt{guided-pomp-review}
skill alongside a meta-skill that enabled it to create a new skill file
upon discovering a reusable workflow or method of analysis during the
review process. This configuration was designed to assess whether AI
could iteratively evolve its own context with minimal human
intervention. Unlike the other three agents which reviewed projects in
parallel, MetaSkill was run sequentially across all 72 projects in
order, so that skill files generated during earlier reviews were
available to inform later ones.

\textbf{Orchestrator}: This agent loaded the same
\texttt{guided-pomp-review} skill file as CourseGuided, but applied it
within a self-contained multi-step pipeline: an initial review, a dual
audit checking evidence grounding and methodological coverage, and a
challenge-judge step that stress-tested each flagged point. Only claims
that survived the challenge process appeared in the final review.
Procedures where an agent orchestrates a multi-stage evaluation process
have been found to improve outcomes for LLM annotations in learning
analytics (Ahtisham et al., 2026).

The POMP review skill was initially drafted by AI, using a prompt to
build a skill summarizing the recommendations of Wheeler et al. (2024).
The course-guided skill was similarly constructed to provide a summary
of the course notes. The eponymous meta-skill skill for the MetaSkill
agent, and the agent description files, were drafted using internet
advice for building self-improving agents. All configuration files were
edited manually during pilot experimentation.

\begin{longtable}[]{@{}cl@{}}
\caption{Classification codes for issues raised in
review}\label{tbl-letter-code}\tabularnewline
\toprule\noalign{}
Code & Description \\
\midrule\noalign{}
\endfirsthead
\toprule\noalign{}
Code & Description \\
\midrule\noalign{}
\endhead
\bottomrule\noalign{}
\endlastfoot
A & AI major finding --- human did not raise \\
B & AI major finding --- human also raised \\
C & AI minor finding --- human did not raise \\
D & AI minor finding --- human also raised \\
E & Human raised --- AI did not address \\
F & Direct contradiction (excluded from analysis) \\
\end{longtable}

Each AI finding was classified into one of the six categories A--F in
Table~\ref{tbl-letter-code}. The classification was carried out by a
team of agents: the Comparator, a HumanExtractor, and the
ComparatorReviewer. The Comparator delegated all reading and analysis to
nested agents rather than reading files itself. A HumanExtractor
instance read the aggregated human review and extracted its distinct
concerns into a structured list, and a new ComparatorReviewer instance
was called once per AI review agent (Baseline, CourseGuided, MetaSkill,
and Orchestrator) to classify the reviewer's findings against the
extracted human list. Each nested agent's output was written to disk
individually, establishing an audit trail, and a deterministic script
assembles the final per-project report from these outputs, while also
identifying consensus misses: human issues classified as Category E for
every one of the four reviewers.

For the quantitative evaluation, we scored each match identified by the
Comparator agent as ``clear,'' ``disputable,'' or ``incorrect.''
Disputable matches were ones where we could see valid logic supporting
the decision, but also lines of reasoning that could lead to alternative
decisions. We found few incorrect matches (7.4\%) and a majority of
clearly correct matches (75\%). See Appendix B for additional details.

Findings were matched to a human issue if they referred to substantially
the same underlying concern, even if phrased differently. The
major/minor distinction (A vs C, B vs D) was determined solely by the AI
reviewer's own label. Each human issue was counted exactly once. A peer
review, carried out by human or AI, that finds some substantive issues
can be helpful whether or not other points are correct, and so our
conclusions should be insensitive to a small fraction of undetected
errors.

The primary quantitative metric was \emph{Human Overlap} defined as
\[\text{Human Overlap} = 1 - \frac{E}{N},\] where \(N\) is the number of
issues raised by the human reviewer for the project. This measures the
fraction of human-confirmed weaknesses independently identified by the
AI agent. Human overlap does not necessarily measure AI review quality
since non-overlapping points may be AI insights missed by humans, or AI
errors, or human errors. It is an asymmetric measure of the similarity
between the AI findings and the human reviews. To assess whether Human
Overlap differed across the four agents, we used the Friedman test
(Friedman, 1937), a non-parametric repeated measures test, treating the
72 projects as repeated units. The null hypothesis (H\_0) is that the
distribution of Human Overlap is the same across all agents, against the
alternative (H\_a) that at least one agent differs. Since the aggregated
human reviews represent instructor-verified review points, Human Overlap
measures the extent to which AI matches validated human assessment. High
Human Overlap is not necessarily desirable if the human reviews are
missing important points. We addressed this concern by examining
Category A findings: major AI findings that no human reviewer raised
(see Section \ref{what-ai-caught-that-humans-missed}). Separately, the
consensus misses at the end of each Comparator report were aggregated
across all 72 projects into a single list. We looked for verbal
descriptions summarizing the types of issues falling into each category.
Frontier AI models can be effective at this topic modeling task (Mu et
al., 2024), and we took advantage of that capability. The full workflow
is detailed in the \texttt{analysis} folder of our code repository.

\section{Results}\label{results}

{

\begin{longtable}[]{@{}lllll@{}}

\caption{\label{tbl-coverage-raw}Human Overlap (\%) by agent and
semester.}

\tabularnewline

\toprule\noalign{}
& W21 & W22 & W24 & W25 \\
Agent & & & & \\
\midrule\noalign{}
\endhead
\bottomrule\noalign{}
\endlastfoot
Baseline & 35.3 & 36.1 & 30.0 & 26.6 \\
CourseGuided & 42.7 & 42.0 & 24.7 & 33.4 \\
MetaSkill & 35.4 & 40.2 & 29.1 & 29.6 \\
Orchestrator & 22.7 & 44.3 & 30.7 & 27.7 \\

\end{longtable}

}

\begin{figure}

\centering{

{\includegraphics[width=14.5cm]{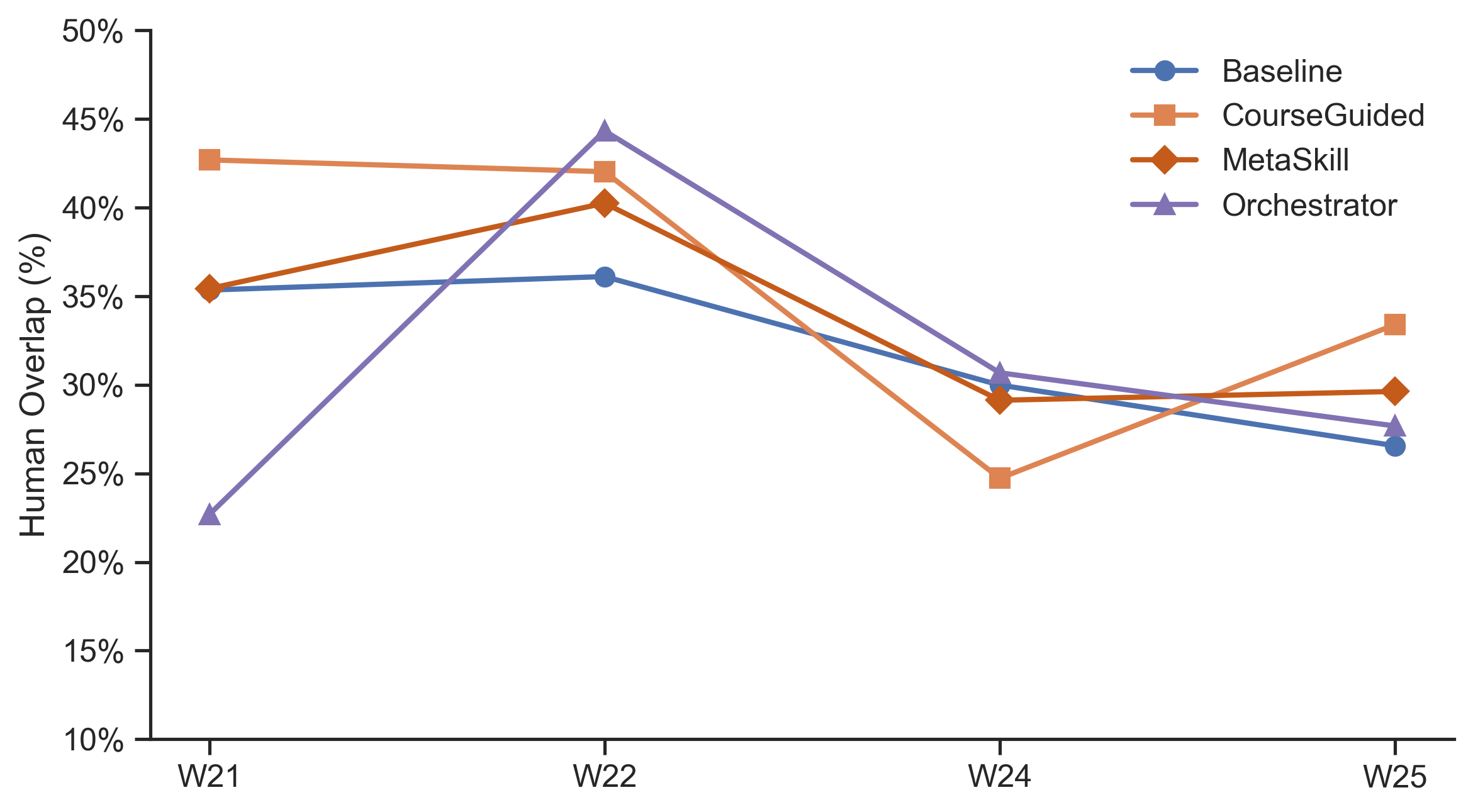}}

}

\caption{\label{fig-coverage-trends}Human Overlap by agent across four
semesters. All agents showed declining coverage in later semesters,
partly due to increasing human reviewer thoroughness.}

\end{figure}

Table~\ref{tbl-coverage-raw} presents the Human Overlap by agent and
semester, computed by averaging each project's own Human Overlap ratio
within that semester. Figure~\ref{fig-coverage-trends} shows the Human
Overlap trend across semesters for each agent. Overall, there is a weak
declining trend, perhaps because human reviews became more thorough over
time: the average number of human-confirmed issues per project grew from
5.3 in W21 to 9.9 in W24 before settling at 9.3 in W25. There are
various other competing hypotheses for the trend: the introduction of a
peer review component in the course might have had an effect on both the
projects and reviews in subsequent years, since students were encouraged
to inspect the publicly posted results from previous years; the COVID-19
pandemic might have affected early years, and the growing capabilities
of AI might have affected later years.

\begin{figure}

\centering{

{\includegraphics[width=14.5cm]{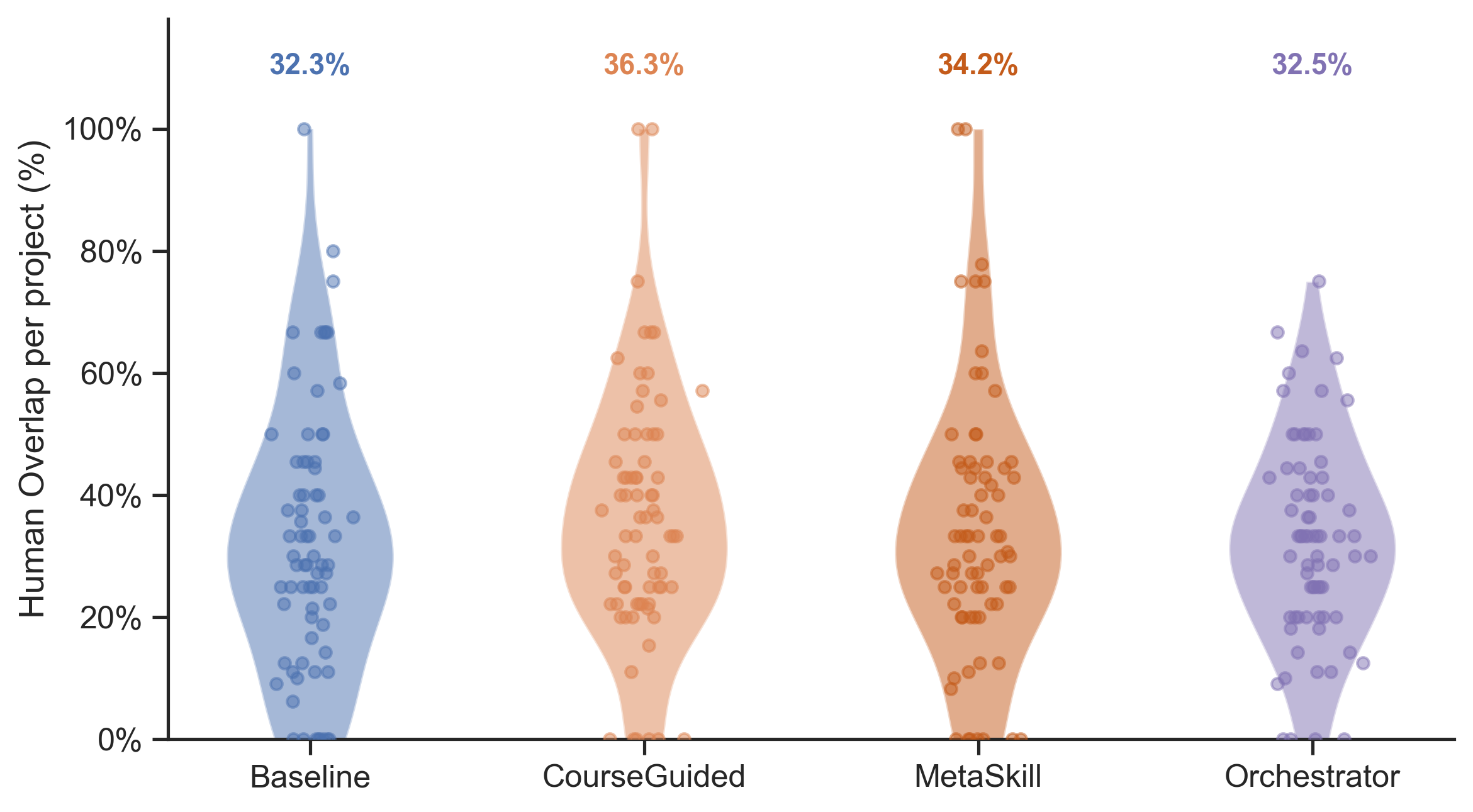}}

}

\caption{\label{fig-per-project}Per-project Human Overlap across all 72
projects for each agent.}

\end{figure}

The Friedman test on Human Overlap across the four agents yielded p =
0.36, suggesting inconclusive evidence against the null hypothesis. The
overall mean overlap rates of 32.3\% (Baseline), 36.3\% (CourseGuided),
34.2\% (MetaSkill), and 32.5\% (Orchestrator) were similar across
agents, with substantial project-level variance visible in
Figure~\ref{fig-per-project}.

A consistent finding across all agents and semesters was that the
majority of each agent's output consisted of AI-unique findings that
human reviewers never raised. The Baseline agent generated approximately
6--7 major findings per project that the human reviewer did not raise,
while the human reviewer raised 5--9 issues per project that no AI
reviewer addressed.

\subsection{What AI Caught That Humans
Missed}\label{what-ai-caught-that-humans-missed}

\label{what-ai-caught-that-humans-missed}

Despite low Human Overlap, AI agents generated a large and qualitatively
consistent set of findings not raised by human reviewers (e.g., case
studies \ref{cs:ai-major-issue}--7 in Appendix A). The findings divided
along the axis of skill file configuration. The Baseline agent's unique
findings frequently required reading and executing code to evaluate,
rather than assessing the statistical argument alone. This required the
AI to understand particle filters, maximum likelihood via the iterated
filtering (IF2) algorithm (Ionides et al., 2015), and implementation via
the pomp R package (King et al., 2016). These technical issues fell into
four recurring categories, described below.

\textbf{Code-level bugs that silently corrupt computation.} The most
consequential class of Baseline-unique findings involved code that ran
without error while computing the wrong quantity. Examples included:
accumulator variables tracking the wrong quantity (e.g., a compartment
in a disease transmission model accumulating recoveries while the
measurement model used it as a proxy for new infections), incorrect
time-step specifications (e.g., \texttt{euler()} instead of
\texttt{discrete\_time()}, introducing spurious sub-steps), and particle
filters inadvertently run on simulated rather than real data---rendering
the reported benchmark log-likelihood meaningless.

\textbf{Data handling and series preparation errors.} The Baseline agent
consistently caught data loaded in reverse chronological order,
frequency parameters miscalibrated to the wrong periodicity (e.g.,
\texttt{frequency=365} for trading-day data with approximately 252
trading days per year), silently dropped observations, and discrepancies
between parameter values stated in text versus implemented in code.

\textbf{Search configuration failures.} The Baseline agent identified
search boxes too narrow to contain the true MLE, search box designs
derived from local search results, and ad hoc parameter choices lacking
statistical justification.

\textbf{Reproducibility failures.} Undefined variables that would cause
runtime errors on re-execution, and hardcoded results in the report that
do not match the output of the submitted code.

In contrast to the baseline agent, the skill-equipped agents
(CourseGuided and MetaSkill) shifted focus toward inference-methodology
completeness, with recurring examples including:

\textbf{Missing or invalid benchmark comparisons.} The CourseGuided
agent raised the absence of non-mechanistic baseline time series model
comparisons (ARIMA, ARMA, GARCH) as a standalone major finding in
approximately 13 of 16 W21 projects. The Baseline agent raised this in
none.

\textbf{Profile likelihood validity.} The skill-equipped agents
identified specific mechanisms that reduced the validity of profile
likelihood analyses. Examples include exclusion of parameters from the
iterated filtering random-walk perturbation specification, maximum
likelihood estimates lying at search box boundaries, and confidence
interval construction based on local rather than global search.

\textbf{Global search initialization anti-pattern.} Both skill-equipped
agents systematically flagged the practice of initializing global
likelihood searches from prior search results. This approach may be an
acceptable practice in some situations, but risks overstatement when
described as a global search. The Baseline agent missed this
anti-pattern in the majority of cases.

\textbf{IF2 configuration completeness.} The skill-equipped agents
audited which parameters were actively perturbed during IF2
optimization, catching cases where key epidemiological parameters were
left unperturbed and therefore could not converge to their maximum
likelihood estimates.

\subsection{What Humans Found That AI Consistently
Missed}\label{what-humans-found-that-ai-consistently-missed}

\begin{figure}

\centering{

{\includegraphics[width=14.5cm]{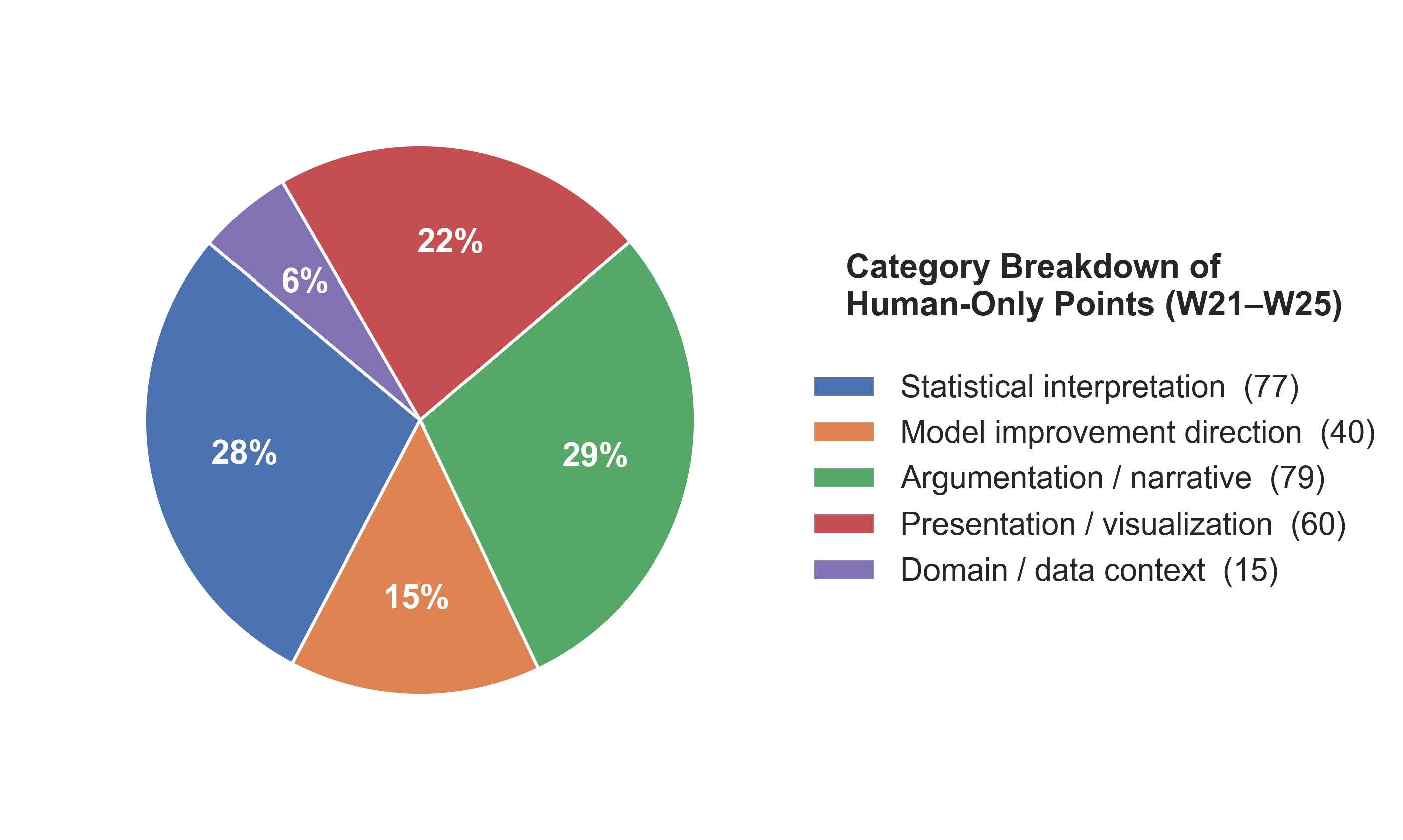}}

}

\caption{\label{fig-themes}Category breakdown of human-only findings
across all semesters (W21--W25). Argumentation/narrative and statistical
interpretation account for the largest, roughly comparable shares,
followed by presentation/visualization.}

\end{figure}

To characterize the structure of human-unique findings, we analyzed 271
human-raised issues that were missed for all four agents. Themes were
identified by having Claude broadly classify each finding by content
into general categories. Appendix A presents individually-verified case
studies illustrating what these categories mean in practice for specific
projects. These clustered into five recurring themes, shown in
Figure~\ref{fig-themes}.

\textbf{Statistical and scientific interpretation (n = 77, 28\%).} Human
reviewers correctly diagnosed the meaning of statistical results in ways
AI consistently failed to replicate. Examples included: identifying flat
profile likelihoods as evidence of non-identifiability rather than
optimization failure, recognizing that bimodal likelihood surfaces
warrant investigation and scientific discussion, and catching logically
inconsistent conclusions such as rejecting a null hypothesis and then
drawing a positive inference from the same p-value. See case study
\ref{cs:statistical-interpretation} in Appendix A

\textbf{Argumentation and narrative coherence (n = 79, 29\%).} Human
reviewers assessed whether the scientific argument was internally
consistent, the research motivation was clearly stated, the model class
was appropriate for the question posed, conclusions followed from the
evidence presented, and the analysis contributed insight beyond prior
work. The AI agents did not engage with these dimensions of review
quality. See \ref{cs:argument-narrative}

\textbf{Presentation and visualization quality (n = 60, 22\%).} Human
reviewers caught specific figure and table problems that required
holistic visual assessment: uninformative or unpolished EDA plots,
reversed time axes in figures, excessive significant figures in tables,
missing captions and figure numbers, and inconsistent or undefined
notation. See \ref{cs:presentation-visualization}

\textbf{Model improvement direction (n = 40, 15\%).} Human reviewers
identified the specific structural change that would resolve an observed
problem. Examples included recognizing that weekly periodicity in COVID
data necessitates a time-varying measurement model or 7-day
differencing, that multiple epidemic waves require additional model
compartments rather than parameter tuning, or that over-dispersed
process noise would improve fit. AI agents identified that models failed
without proposing actionable structural remedies.

\textbf{Domain and data context (n = 15, 6\%).} Human reviewers applied
domain knowledge that went beyond the code. Examples included
questioning how a dataset operationalized contested political concepts
(e.g., whether an election was ``free and fair''), identifying when
stock data used unadjusted close prices affected by splits, and
recognizing when modeling assumptions conflicted with domain-specific
data conventions.

\subsection{Skill File Focus Shift}\label{skill-file-focus-shift}

\begin{figure}

\centering{

{\includegraphics[width=14.5cm]{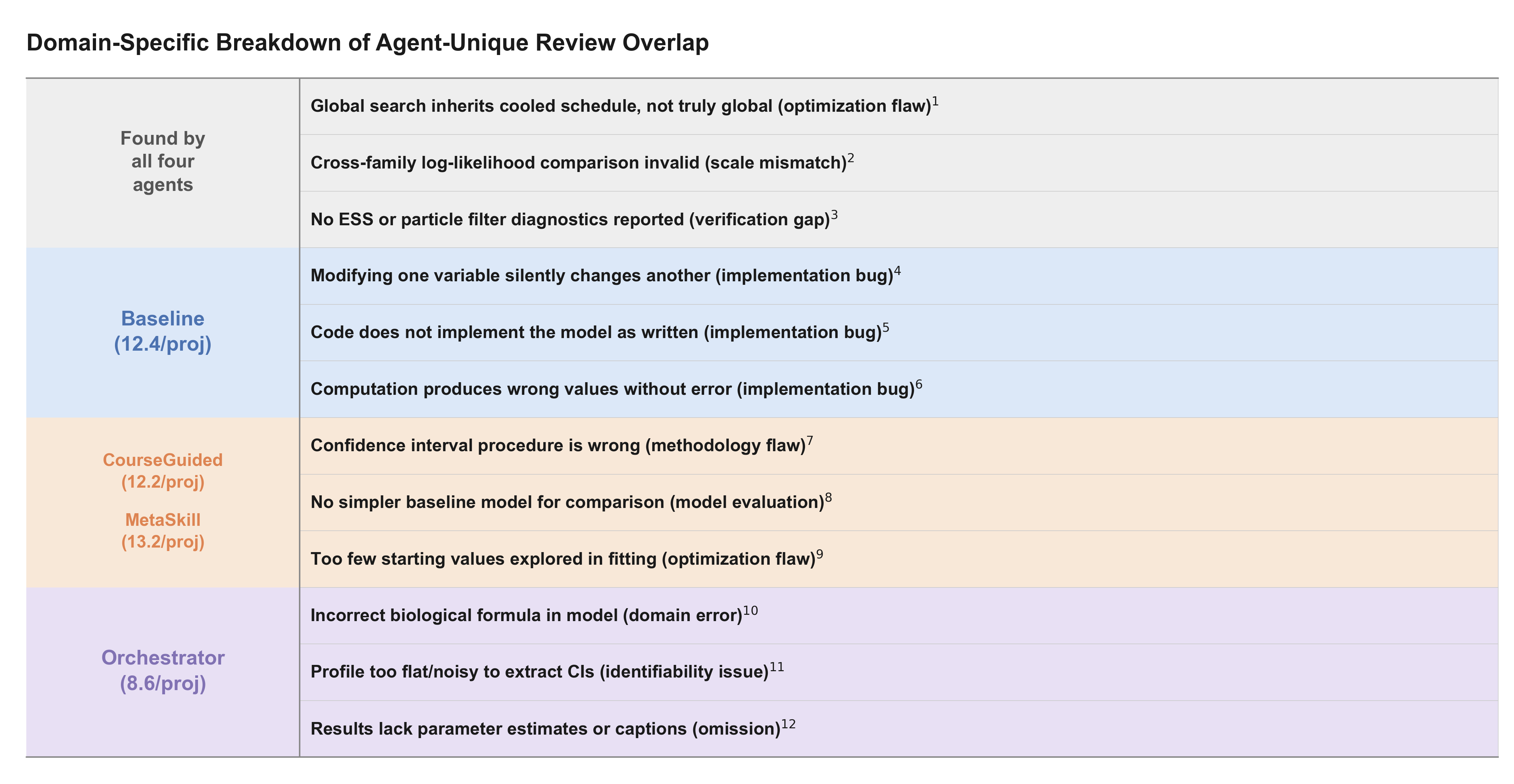}}

}

\caption{\label{fig-matrix}Domain-specific comparison of issues raised
by AI but not by human reviewers, aggregated across all semesters. The
Baseline agent performed best at implementation checks; skill-equipped
agents redirected focus to inference methodology.}

\end{figure}

The lack of a statistically detectable difference in Human Overlap
across agents masked a qualitatively significant divergence in the
\emph{types} of findings each agent generated. Figure~\ref{fig-matrix}
illustrates this divergence: the Baseline agent captured implementation
errors at consistently high rates, while skill-equipped agents
systematically flagged inference-methodology violations that the
Baseline missed.

This trade-off was directional and category-specific. The Baseline agent
focused on data, code, and reporting errors: particle filter object
identity, silent computation corruption, text-code mismatches at the
parameter level, and attribution failures. For example, see
\ref{cs:baseline-major-error} in Appendix A. The skill-equipped agents
focused on inference-methodology completeness: benchmark comparison
requirements, profile likelihood validity, global search initialization
correctness, and model diagnostic outputs, following the context framed
by the skill files. An example of this is
\ref{cs:courseguided-metaskill-benchmark} in Appendix A, where both
agents with the relevant skill files flagged this, while Baseline and
Orchestrator did not. Orchestrator's unique findings took a different
form, including identifiability concerns that no other agent raised; see
\ref{cs:orchestrator-sigma-nu-boundary} in Appendix A. We read this as
evidence that a skill file's domain-specific instructions redirect an
agent's attention toward a particular class of issue. An issue within
that domain can produce a specific catch, as above, but this redirection
can come at the expense of more basic, project-specific bugs outside
that domain. This is illustrated by a case study for the Baseline agent
(\ref{cs:baseline-major-error}) where a process-model bug was missed by
every skill-equipped agent, but caught by the unconstrained Baseline
agent. This is consistent with the figure's own construction, where its
rows were derived by asking a separate AI instance to identify the most
commonly recurring themes among each agent's unique findings, a
methodology that by construction surfaces frequently-repeated patterns
rather than rare, project-specific critical bugs.

The trade-off is consistent with the Friedman test result: the skill
files redirected the agent to a different category of points without
increasing total overlap with the human-identified weaknesses.
CourseGuided and MetaSkill are grouped together (and their totals
averaged) in Figure~\ref{fig-matrix} because their qualitative finding
patterns were nearly identical despite MetaSkill accumulating additional
skill files throughout its sequential run. This similarity reflects a
point of diminishing returns: beyond a certain point, adding context
through skill files did not produce meaningfully different coverage.
This coverage-level similarity concerns which categories of issue were
flagged, not the rigor of the individual diagnosis:
\ref{cs:courseguided-metaskill-benchmark} and
\ref{cs:dN_RS-wrong-compartment} in Appendix A shows the two agents
flagging the same category of concern with markedly different rigor.

\section{Discussion}\label{discussion}

Our results showed that the AI reviewers exhibited a jagged capability
profile in POMP peer review. The findings can be organized around three
central observations.

\textbf{1. AI and human reviewers largely detected different problems.}
Across all four semesters and all agents, fewer than 40\% of
human-confirmed weaknesses were independently identified by any AI
reviewer. The majority of AI findings were entirely absent from human
reviews, while most human-identified issues went unaddressed by AI. This
complementary structure held regardless of skill file configuration. AI
reviewers can serve as a valuable supplement to human review precisely
because they cover ground that human reviewers---who typically do not
read and execute source code---would not otherwise check. Appendix A
presents four such examples in detail: substantial bugs that we
independently verified, each identified by all four agent configurations
yet missed by the human reviewers
(\ref{cs:ai-major-issue}--\ref{cs:poor-mans-profile-likelihood})

\textbf{2. The addition of skill files tuned jaggedness rather than
resolving it.} When equipped with a domain-specific skill file, the
agent's focus pivoted toward inference methodology within the context
detailed therein---producing stronger coverage of benchmark comparisons,
profile likelihoods, convergence diagnostics, and POMP-specific
inference standards. The agent's assigned skills also affected the
classification of a given issue as major or minor. However, the skills
did not significantly affect the overlap with the human reviewers, and
therefore did not fundamentally affect the jaggedness of the review.
This suggests that scientific review may be in the class of tasks for
which the performance of modern AI is relatively insensitive to the
wording of the instructions Wang et al. (2026). Beyond this aggregate
pattern, \ref{cs:courseguided-metaskill-benchmark} in Appendix A shows
this tuning effect at the level of a single finding. CourseGuided and
MetaSkill both caught an invalid benchmark comparison that Baseline and
Orchestrator missed entirely, but with markedly different rigor
traceable to the specific content of each agent's skill file. This may
help explain why the skills did not move the overlap with human
reviewers. A skill file can help target a genuine issue and sharpen an
agent's rigor within the specific domain it targets, but does not grant
flexibility beyond its pre-specified domain.

\textbf{3. The human-unique layer was characterized by judgment and
context that the AI agents did not supply.} The five recurring
categories of human-unique findings (statistical interpretation,
argumentation, presentation quality, model improvement direction, and
domain context) shared a common feature: they required integrating
information across the project holistically, applying domain knowledge
that went beyond the text, and exercising scientific judgment about what
questions and answers are worthwhile. Across the different skill
contexts, these AI agents did not raise these issues. For example, the
AI reviewers could verify whether a profile likelihood was correctly
computed, but not whether the interpretation of a flat profile was
scientifically sound. A flat profile likelihood suggests a lack of
marginal information about a parameter, and this could be problematic,
or could be just a fact of marginal relevance if other interesting
parameters show curvature in their respective profile likelihoods, or
could suggest reparameterizing to look for a well-identified parameter
combination. Some diligent human referees were able to address these
interpretive issues. As another example, the AI agents could detect a
missing benchmark comparison, but not whether the model class was
fundamentally inappropriate for the scientific question.

Our results cannot necessarily be extrapolated to other agent
configurations or other foundation models. The positive finding, that
our AI review agents identified useful insights consistently missed by
humans, is robust to this limitation. The negative finding, that our AI
review agents had significant limitations, could be reassessed by future
research. Such limitations will exist as long as AI capabilities remain
jagged, but the boundary of the limitations will change. A
game-theoretic analysis of jaggedness suggests that it places high value
on understanding the specific capabilities and limitations of AI (Gans,
2026), advocating for the value of benchmarking studies such as this.

\section{Conclusions}\label{conclusions}

We evaluated four Claude agents with varying skill file configurations
on 72 POMP analysis projects, comparing their output to open-access
human peer reviews. The central finding was that AI reviewers exhibited
a jagged capability profile: they proficiently caught technical
implementation errors and invalid inference methodology that human
reviewers often overlooked, while consistently failing to match human
standards in statistical interpretation, narrative coherence, model
improvement direction, and domain-informed critique. The addition of
domain-specific guidance, via skill file configurations, shifted the
agent's focus without substantially affecting jaggedness, leaving the
overall proportion of human-confirmed weaknesses detected statistically
indistinguishable across all four agent configurations.

The complementarity between AI and human review, with each detecting a
largely non-overlapping set of problems, suggests that AI review can
currently serve as a valuable supplement to human review but not as a
replacement for human judgment. AI is particularly well-suited to
catching implementation-layer errors that human reviewers may not have
the time to identify. The use of AI to develop scientific data analysis
will hasten the use of complex methodology that cannot readily be fully
evaluated without AI assistance.

We studied graduate student peer review of course final projects, which
is a more controlled environment than journal peer review. In
particular, students were required to submit source code that reproduced
the project report, using R markdown. The AI agents may have benefitted
more than the humans from this high expectation for reproducibility.
Another difference is that final projects may have more mistakes, since
they represent around two weeks of work. However, we postulate that
mistakes made by statistics graduate students after taking an advanced
course may be representative of mistakes made by practitioners using
similar statistical methods for scientific investigations.

Improvements in AI technology, and/or improvements in implementation of
the peer review agent instructions, will presumably lead to improvements
in AI review capabilities. Our experiment documents AI capabilities at a
particular time, at a particular task. If these results hold more
generally, they would imply that scientific journals should accept a
role for AI in peer review.

\subsection*{Acknowledgments}\label{acknowledgments}
\addcontentsline{toc}{subsection}{Acknowledgments}

The authors thank Aaron Abkemeier, Fred Feinberg and Bo Yang for helpful
suggestions.

\textbf{Ethics}. This study analyzes publicly available and
de-identified data, and was therefore determined not to require human
subject regulation by the University of Michigan Institutional Review
Board (HUM00293579).

\textbf{Use of AI}. AI tools were used as described in Materials and
Methods. Claude Code Opus 4.7 and Chat GPT 5.5 were used to provide
review of this manuscript, and we are grateful for their helpful
suggestions.

\textbf{Disclosure}. The authors declare no competing interests. This
work was supported by National Science Foundation grant OAC-2607869.

\textbf{Data availability} The data, source code, and results were
archived on Zenodo in addition to the GitHub links provided in the text.
The corresponding digital object identifiers are
\href{https://doi.org/10.5281/zenodo.22997697}{10.5281/zenodo.22997697}
(for W21),
\href{https://doi.org/10.5281/zenodo.22997783}{10.5281/zenodo.22997783}
(for W22),
\href{https://doi.org/10.5281/zenodo.22997878}{10.5281/zenodo.22997878}
(for W24),
\href{https://doi.org/10.5281/zenodo.22997892}{10.5281/zenodo.22997892}
(for W25),
\href{https://doi.org/10.5281/zenodo.22998009}{10.5281/zenodo.22998009}
(for full records of the source code and results).

\subsection*{Bibliography}\label{bibliography}
\addcontentsline{toc}{subsection}{Bibliography}

\protect\phantomsection\label{refs}

\section*{Appendix A. Case Studies}\label{sec-appendix-cs}
\addcontentsline{toc}{section}{Appendix A. Case Studies}

\footnotesize

These vignettes illustrate the scientific and statistical issues arising
in the project reports, human reviews and AI reviews, as well as the
categorizations of these issues in our data analysis. Of necessity, we
present only selected items from the full dataset available online.

\refstepcounter{csnum}\label{cs:ai-major-issue}

\textbf{CS\arabic{csnum}.} \textbf{An example where the Comparator agent
determined that AI reviewer agents had found a major issue missed by the
humans} \textbf{(W21, Project 1)}. All four review agents assessed that
setting \texttt{H=I} in the measurement model was a major error. Here,
\texttt{I} is a computational variable matching the model quantity
\(I(t)\), the continuous-time latent instantaneous number of infected
individuals. \texttt{H} is a so-called accumulator variable linking
\texttt{I} to observed data via the measurement model. Typically,
\texttt{H} is set to be the number of new infections during a reporting
interval, or perhaps equal to the number of removed infections if
detected cases are quarantined. The measurement model also includes a
reporting rate, and in this project it has a binomial specification,
coded as \texttt{lik = dbinom(reports,H,rho,give\_log)}. The measurement
model is only explicitly defined in the source code, since some code is
suppressed for readability. The \texttt{H=I} error is apparent in both
the report and its source code, but was missed by the human referees.
The human review did notice some issues that resulted from the problem,
e.g., human review point \#2 included the comment, ``perhaps the noise
modeling in the process and/or measurement model is a misfit,'' but
noticing evidence of a problem is less insightful than identifying the
problem itself. The comparator assessed human point \#2 to be ``human
only'' and the \texttt{H=I} issue as ``AI only.'' This is not an
unreasonable decision, but it exemplifies the difficulty in making a
binary decision about whether issues match each other. The human
approach, suspecting a problem even if it is not identified, can be key
to error-correcting on long and complex tasks. The phenomenon that
humans make more errors, but are better at recovering after an error,
has been proposed as a general explanation of the jagged intelligence
phenomenon that current AI can be more accurate at individual tasks and
yet show sub-human fragility at some complex tasks (Mineault et al.,
2026).

\refstepcounter{csnum}\label{cs:dmeasure-rbinom-bug}

\textbf{CS\arabic{csnum}.} \textbf{A fundamental measurement-model
violation found independently by all four agents and missed by the
humans} \textbf{(W22, Project 3)}. A POMP model has two parts: a
hidden/latent process that is not observed directly, and a measurement
model describing the actual data as a noisy or incomplete observation of
the hidden process. The function \texttt{dmeasure} computes the density
\(p(y_n | x_n, \theta)\), using this to weight each simulated
hypothetical trajectory of the hidden state by how well it explains the
real data. This function should be deterministic, given its inputs.
However, this project's \texttt{dmeasure} instead begins by drawing a
fresh random value on every call,
\texttt{Views = rbinom(N-S, 1-exp(-Beta*S/N))}. This same variable,
\texttt{Views}, is used in constructing the mean and standard deviation
of the reported likelihood. Since \texttt{Views} is being regenerated
independently each time \texttt{dmeasure} is evaluated, the same state
and parameters yield a different log-likelihood value on every call.
This violation of the formal POMP model assumptions risks invalidating
theoretical guarantees for POMP methods such as the particle filter. The
severity of this violation is unclear, but it should be noted.

\refstepcounter{csnum}\label{cs:dN_RS-wrong-compartment}

\textbf{CS\arabic{csnum}.} \textbf{A process-model coding bug found
independently by all four agents, where only two of four diagnosed a
related flaw} \textbf{(W22, Project 4)}. A compartmental model is a POMP
model describing a population for which each individual falls in one
compartment. Such models are specified by the transitions rates between
compartments. This project concerns an epidemiological compartment model
for which the total rate of transition from a recovered state to a
susceptible state (i.e., loss of immunity) was mathematically specified
to depend on the number of recovered individuals. However, the code
computes this transition as
\texttt{dN\_RS = rbinom(I, 1-exp(-mu\_RS*dt))}, with a rate dependent on
the infected compartment, \(\mathrm{I}\). This coding error was
discovered by all four AI agents. There is another related bug here,
that only two of the four agents flagged: the transition equation
\texttt{R += dN\_PR + dN\_IR} does not include the decrement for loss of
immunity. Among other problems, this violates the conservation of the
total population, as individuals added to \texttt{S} are not subtracted
from \texttt{R}. The Baseline and CourseGuided agents both identified
this second bug explicitly, alongside the first. The Orchestrator agent
did not address the second issue and provided a fix for only the first
issue, which would not preserve conservation. The MetaSkill agent
proposed an incorrect solution: \texttt{R -= dN\_RS + dN\_PR + dN\_IR}.
The Baseline agent saw deepest into the details of the code, perhaps
because it was least burdened by additional instructions on how to
conduct the review. The CourseGuided agent's finding, by contrast, maps
directly onto a specific instruction in its \texttt{531\_references}
skill file, which lists checking for a ``Compartment model that violates
conservation of individuals (compartments don't sum to population)'' as
a genuine review point. This finding is an example of how
domain-specific skill files can sharpen an agent's rigor within their
specific domain of coverage but confer little advantage outside it, as
evidenced here by MetaSkill and Orchestrator, neither of which possessed
\texttt{531\_references}.

\refstepcounter{csnum}\label{cs:poor-mans-profile-likelihood}

\textbf{CS\arabic{csnum}.} \textbf{Identification of a methodological
error in profile likelihood calculation} \textbf{(W24, Project 16)}. All
four AI review agents were concerned about confidence intervals that
claimed to be constructed via profile likelihood. For example, according
to the Baseline agent, ``The figures presented for \texttt{Beta\_v},
\texttt{Beta\_u}, \texttt{mu\_IR\_v}, \texttt{mu\_IR\_u}, and the two
ratios are produced by filtering the global search results and plotting
the top values. A profile likelihood requires, for each fixed value of
the parameter of interest, maximizing the likelihood over all other free
parameters. The plots shown are ad hoc slices of the global search
cloud, which gives a much wider and less informative picture than a true
profile. Confidence intervals derived from these plots using the
chi-squared cutoff are not statistically valid.'' It is technically
correct that the envelope of the scatterplot of Monte Carlo searches
presented is not a profile likelihood. This graph always lies under the
profile, approaching the profile as the number of Monte Carlo searches
increases. The plot is sometimes called a ``poor man's profile,'' and it
can be a useful preliminary approximation to a profile for exploratory
analysis, since it can be considerable additional work to compute
profile likelihoods for all parameters. The agents correctly identified
the error in calling it a profile, though they overestimate the
magnitude of the flaw. The humans, by contrast, failed to note the
discrepancy.

\refstepcounter{csnum}\label{cs:baseline-major-error}

\textbf{CS\arabic{csnum}.} \textbf{A major error reported by the
Baseline agent but not the other agents or the humans} \textbf{(W25,
Project 5)}. This concerns the following technical comment made by the
Baseline agent:

\vspace{1.2mm}

\hspace{4mm}\parbox{5.6in}{
The code in chunk \texttt{setup-immi} updates the R objects \texttt{paramnames} and \texttt{rproc} in memory, and sets new coefficients on \texttt{seir\_spline\_model} via \texttt{coef(seir\_spline\_model) <- c(...)}.
However, there is no second \texttt{pomp(...)} call that rebuilds the model with the new \texttt{rproc} (containing \texttt{immigration\_rate}) and updated \texttt{paramnames}.
The original \texttt{seir\_spline\_model} was compiled against the old C-snippet and old \texttt{paramnames}.
Setting coefficients does not recompile the process model.
As a result, the "POMP Model with Immigration" local and global searches almost certainly ran using the original process model without immigration dynamics, invalidating the entire comparison between the two POMP models. This is the most serious technical error in the project.
}

\vspace{1.2mm}

This AI review comment is supported by subsequent human investigation of
the code, specifically by inspection of lines 425 and 565 of
\texttt{blinded.Rmd}. In this case, the AI agent least burdened by
additional instructions saw deepest into the details of the code. This
is an example of how providing domain-specific skill files can sometimes
confer little to no advantage. This finding illustrates the specific
category in Figure~\ref{fig-matrix} where the code does not implement
the model as written, one of the characteristic features associated with
Baseline agent's unique findings.

\refstepcounter{csnum}\label{cs:courseguided-metaskill-benchmark}

\textbf{CS\arabic{csnum}.} \textbf{A finding raised by the CourseGuided
and MetaSkill agents but not the Baseline or Orchestrator agents}
\textbf{(W21, Project 8)}. This project fits three basic time series
models (ARMA/ARIMA, GARCH, ARMA-GARCH) to provide benchmarks for
comparison with the proposed mechanistic model. The benchmarks computed
an exact Gaussian likelihood (or conditional likelihood, for GARCH) on
the raw data scale. Later in the project, four POMP models (Gaussian
HMM, Student-t HMM, Gaussian HMM AR(1), and Heston) were constructed and
fitted. These have a stochastic hidden state that cannot be integrated
out in closed form, so their likelihood was approximated via
\texttt{pomp}`s particle filter. The project builds all seven models'
log-likelihoods and AIC values into one combined table, and concludes
from it that the Heston model yields the best performance in terms of
likelihood. Both the CourseGuided and MetaSkill agents flagged this
comparison as invalid, but with different lines of reasoning.
CourseGuided identifies that the GARCH(1,1) model was fit on the
demeaned series (\texttt{ts - mean(ts)}) while every other model in the
table was fit on the raw series. It also flags that the \texttt{tseries}
package's GARCH likelihood function reports non-standard values, which
reflects an item documented in the course's own list of common errors
(\texttt{531\_references}, Error 2.9). This is consistent with the same
skill file's more general instruction that AIC comparisons across ARIMA
and POMP models require verifying a common data representation and
likelihood definition (Error 2.2), rather than being invalid by default
and it warns against overgeneralizing this into a named misconception:
claiming likelihoods from different model classes are not comparable
(Error 1.14). MetaSkill reached a similar headline conclusion, writing
that ``the conclusion that the Heston model is the `best' model based on
the highest log-likelihood is valid only within the POMP-family models
that share a comparable observation model structure.'' This reasoning is
not correct and is the same kind of overgeneralization that Error 1.14
warned against. This broad heuristic may be traced back to a skill file
the MetaSkill agent generated for itself in an earlier run and reused
here: \texttt{sarima-baseline-audit} states in its procedure: ``if the
observation models differ, flag the comparison as invalid.'' Neither the
Baseline nor the Orchestrator agent raised this concern. This project
has an unusually brief human review that raises only one weakness and
does not address this point.

\refstepcounter{csnum}\label{cs:orchestrator-sigma-nu-boundary}

\textbf{CS\arabic{csnum}.} \textbf{A finding raised only by the
Orchestrator agent} \textbf{(W22, Project 1)}. This project models the
growth rate of online game player counts as a financial asset return,
fitting a stochastic-leverage volatility model. The latent process (G)
evolves as a Gaussian random walk with standard deviation parameter
\texttt{sigma\_nu}. The report's table of maximum likelihood parameters
shows an estimate \texttt{sigma\_nu = 0}, for which the process becomes
a fixed constant for the entire series and the leverage,
\(R_n = \tanh(G_n)\), is deterministic. The Orchestrator agent's review
classifies this finding as an identifiability issue, recommending that a
profile likelihood be computed across a range of \texttt{sigma\_nu}
values to check whether the parameter is actually identifiable. The
project's Conclusion section does not acknowledge this nor the
implications for the model's core stochastic-leverage claim. The
Baseline, CourseGuided, and MetaSkill agents did not identify this
parameter collapse. No human review point flagged it. This finding is
representative of the identifiability-related category that
Figure~\ref{fig-matrix} associates with the Orchestrator agent's unique
findings.

\refstepcounter{csnum}\label{cs:domain-data-context}

\textbf{CS\arabic{csnum}.} \textbf{Two human-only points categorized as
``Domain/data context.''} \textbf{(i) W21, Project 10}. There is an
erroneous instantaneous drop in susceptibles evident when this is
plotted. It later becomes clear that this corresponds to a huge spike in
cases. A referee noticed the following: ``I try to search for the news
about this abnormality but cannot get a clear answer. Instead, I find a
piece of news that could help to explain it. According to FOX 5 Atlanta
Digital Team, `The spike in cases is also a spike in reported tests. The
Georgia Department of Public Health said about 12,000 of the cases
reported today are all from one facility in metro Atlanta that is
catching up on reporting its numbers for the week.'\,'' \textbf{(ii)
W24, Project 1}. The dataset must somehow determine whether an election
is free and fair, since many autocracies claim to have a democratic
mandate. A brief explanation would help the reader.

\refstepcounter{csnum}\label{cs:statistical-interpretation}

\textbf{CS\arabic{csnum}.} \textbf{Two human-only points categorized as
``Statistical interpretation.''} \textbf{(i) W21, Project 7}. The
project report wrote, ``Profile is flat, suggesting convergence issues
or nonlinearity of the likelihood surface.'' Neither of these is
necessary for a flat profile. There can be a linear tradeoff between
parameters, and perfect convergence, and a flat profile will result. An
appropriate conclusion could be weak identifiability, or
non-identifiability for a perfectly flat profile. \textbf{(ii) W22,
Project 2}. The global search suggests a bimodality: a collection of
searches seem to reach a different and inferior region of parameter
space. This is not scientifically critical - the second mode has much
lower likelihood - but may help to understand numerical issues.

\refstepcounter{csnum}\label{cs:presentation-visualization}

\textbf{CS\arabic{csnum}.} \textbf{Two human-only points categorized as
``Presentation/visualization''}. \textbf{(i) W22, Project 4}. The plot
titled ``differenced data'' is differenced log data, but the ACF next to
it is unlogged data. None of this is apparent unless you study the
source code. \textbf{(ii) W22, Project 18}. Where possible, numbers
should not be hard-coded in the Rmd document. Rather, they should be
referenced using inline R expressions.

\refstepcounter{csnum}\label{cs:argument-narrative}

\textbf{CS\arabic{csnum}.} \textbf{Two human-only points categorized as
``Argumentation/Narrative.''} \textbf{(i) W22, Project 13}. For the
introduction, it is better to include the background (epidemic
situation) in the researched regions, California and Texas. \textbf{(ii)
W24, Project 2}. The description of what is denoted by
\texttt{peak\_rodent\_year} was a bit lacking. It was only described as,
``Peak rodent year is scored as `yes', otherwise `no','' but it does not
describe what is meant by that or how it is decided or when.

\normalsize

\section*{Appendix B. Comparator
Validation}\label{sec-appendix-validation}
\addcontentsline{toc}{section}{Appendix B. Comparator Validation}

Table Table~\ref{tbl-validation-counts} summarizes the findings of the
Comparator validation exercise by agent and semester. 5 projects were
selected at random from each semester, where each Comparator-match was
manually read and scored as a ``clear match,'' ``disputable match,'' or
``incorrect match.'' The selected projects were divided by AI agent
among two independent co-authors. The proportion of incorrect matches
was robust to this decision, as evident by the low frequency of
incorrect matches across all agent-semester combinations.

\begin{longtable}[]{@{}lllll@{}}

\caption{\label{tbl-validation-counts}Comparator match-decision score
(Clear Match, Disputable Match, Incorrect Match) by agent and semester.}

\tabularnewline

\caption{}\label{T_2b6d2}\tabularnewline
\toprule\noalign{}
Semester & Agent & Clear Match & Disputable Match & Incorrect Match \\
\midrule\noalign{}
\endfirsthead
\toprule\noalign{}
Semester & Agent & Clear Match & Disputable Match & Incorrect Match \\
\midrule\noalign{}
\endhead
\bottomrule\noalign{}
\endlastfoot
W21 & Baseline & 4 & 1 & 2 \\
W21 & CourseGuided & 6 & 7 & 2 \\
W21 & MetaSkill & 6 & 3 & 0 \\
W21 & Orchestrator & 5 & 2 & 0 \\
W22 & Baseline & 11 & 3 & 1 \\
W22 & CourseGuided & 18 & 1 & 0 \\
W22 & MetaSkill & 13 & 0 & 0 \\
W22 & Orchestrator & 12 & 0 & 0 \\
W24 & Baseline & 14 & 4 & 2 \\
W24 & CourseGuided & 12 & 3 & 3 \\
W24 & MetaSkill & 18 & 3 & 1 \\
W24 & Orchestrator & 10 & 4 & 4 \\
W25 & Baseline & 11 & 1 & 0 \\
W25 & CourseGuided & 9 & 7 & 2 \\
W25 & MetaSkill & 12 & 1 & 0 \\
W25 & Orchestrator & 12 & 1 & 0 \\

\end{longtable}

\section*{Appendix C. Supplementary Material}\label{sec-supp}
\addcontentsline{toc}{section}{Appendix C. Supplementary Material}

Tables Table~\ref{tbl-raw-counts} and Table~\ref{tbl-themes} summarize
the full review output of the agents and humans. The complete
experimental results are available online (see Materials and Methods).

\begin{longtable}[]{@{}llllllll@{}}

\caption{\label{tbl-raw-counts}Raw finding counts (A, B, C, D, E, F) by
agent and semester.}

\tabularnewline

\caption{}\label{T_406a6}\tabularnewline
\toprule\noalign{}
Semester & Agent & A & B & C & D & E & F \\
\midrule\noalign{}
\endfirsthead
\toprule\noalign{}
Semester & Agent & A & B & C & D & E & F \\
\midrule\noalign{}
\endhead
\bottomrule\noalign{}
\endlastfoot
W21 & Baseline & 91 & 22 & 114 & 12 & 58 & 1 \\
W21 & CourseGuided & 94 & 22 & 115 & 11 & 53 & 0 \\
W21 & MetaSkill & 107 & 21 & 113 & 10 & 54 & 2 \\
W21 & Orchestrator & 60 & 13 & 95 & 9 & 63 & 2 \\
W22 & Baseline & 139 & 37 & 137 & 29 & 115 & 3 \\
W22 & CourseGuided & 130 & 49 & 139 & 25 & 105 & 2 \\
W22 & MetaSkill & 148 & 45 & 147 & 26 & 109 & 2 \\
W22 & Orchestrator & 83 & 39 & 100 & 34 & 103 & 3 \\
W24 & Baseline & 92 & 27 & 105 & 12 & 108 & 5 \\
W24 & CourseGuided & 78 & 24 & 119 & 13 & 112 & 8 \\
W24 & MetaSkill & 92 & 37 & 114 & 18 & 105 & 4 \\
W24 & Orchestrator & 48 & 34 & 79 & 13 & 108 & 1 \\
W25 & Baseline & 107 & 21 & 110 & 15 & 116 & 2 \\
W25 & CourseGuided & 88 & 27 & 114 & 23 & 109 & 1 \\
W25 & MetaSkill & 99 & 29 & 129 & 12 & 114 & 2 \\
W25 & Orchestrator & 72 & 22 & 83 & 16 & 114 & 2 \\

\end{longtable}

\begin{longtable}[]{@{}lll@{}}

\caption{\label{tbl-themes}Counts of human-only findings (E category for
all agents) by theme across all semesters.}

\tabularnewline

\caption{}\label{T_9d88d}\tabularnewline
\toprule\noalign{}
Theme & Count & Percentage (\%) \\
\midrule\noalign{}
\endfirsthead
\toprule\noalign{}
Theme & Count & Percentage (\%) \\
\midrule\noalign{}
\endhead
\bottomrule\noalign{}
\endlastfoot
Statistical interpretation & 77 & 28.4 \\
Model improvement direction & 40 & 14.8 \\
Argumentation / narrative & 79 & 29.2 \\
Presentation / visualization & 60 & 22.1 \\
Domain / data context & 15 & 5.5 \\

\end{longtable}

\end{document}